%% file: main.tex
\documentclass[journal]{IEEEtran}

\ifCLASSINFOpdf
\else
\fi
\usepackage{graphicx}
\usepackage{color}
\usepackage{amsmath,amssymb,amsfonts}
\usepackage[colorlinks,
            linkcolor=blue,
            anchorcolor=black,
            citecolor=green]{hyperref}

\newcommand{\minew}[1]{{\color{black}{#1}}}
\newcommand{\miold}[1]{\iffalse{#1}\fi}

\begin{document}

\title{Autonomous Crowdsensing: Operating and Organizing Crowdsensing for Sensing Automation}

\author{
        Wansen~Wu$^{\dagger}$,\IEEEmembership{}
        Weiyi~Yang$^{\dagger}$,\IEEEmembership{}
        Juanjuan Li, \IEEEmembership{~Member,~IEEE,}
        Yong~Zhao,\IEEEmembership{}
        Zhengqiu~Zhu,\IEEEmembership{}
        Bin~Chen,\IEEEmembership{}
        Sihang~Qiu,\IEEEmembership{}
        Yong Peng*,\IEEEmembership{}
        and Fei-Yue~Wang*, \IEEEmembership{~Fellow,~IEEE}
\thanks{Manuscript received December XX, 2023; revised December XX, 2023; accepted December XX, 2023. This study is supported by the National Natural Science Foundation of China (62202477, 62173337, 21808181, 72071207), and Youth Independent Innovation Foundation of NUDT (ZK-2023-21). (\textit{Wansen Wu and Weiyi Yang contributed equally to this work}.)(\textit{Corresponding author: Yong Peng and Fei-Yue Wang}).}
\thanks{Wansen Wu, Weiyi Yang, Yong Zhao, Zhengqiu Zhu, Bin Chen, and Yong Peng are with the College of Systems Engineering, National University of Defense Technology, Changsha 410073, Hunan Province, China; and also with Hunan Institute of Advanced Technology, Changsha 410073, Hunan Province, China. 
(email:~wuwansen14@nudt.edu.cn;~yangweiyi15@nudt.edu.cn;~zhaoyong15@nudt.edu.cn;zhuzhengqiu12@nudt.edu.cn;chenbin06@nudt.edu.cn;~sihangq@acm.org;~yongpeng@nudt.eud.cn).}
\thanks{
Juanjuan Li is with the State Key Laboratory of Multimodal Artificial Intelligence Systems, Institute of Automation, Chinese Academy of Sciences, Beijing 100190, China. (email:~juanjuan.li@ia.ac.cn).
}
\thanks{Fei-Yue Wang is with the State Key Laboratory for Management and Control of Complex Systems, Institute of Automation, Chinese Academy of Sciences, Beijing 100190, China. 
(email:~feiyue@ieee.org).}

}

\markboth{Journal of \LaTeX\ Class Files,~Vol.~XX, No.~XX, December~2023}%
{Shell \MakeLowercase{\textit{et al.}}: Bare Demo of IEEEtran.cls for IEEE Journals}

\maketitle

\begin{abstract}
The precise characterization and modeling of Cyber-Physical-Social Systems (CPSS) requires more comprehensive and accurate data, which imposes heightened demands on intelligent sensing capabilities. To address this issue, Crowdsensing Intelligence (CSI) has been proposed to collect data from CPSS by harnessing the collective intelligence of a diverse workforce. Our first and second Distributed/Decentralized Hybrid Workshop on Crowdsensing Intelligence (DHW-CSI) have focused on principles and high-level processes of organizing and operating CSI, as well as the participants, methods, and stages involved in CSI. This letter reports the outcomes of the latest DHW-CSI, focusing on Autonomous Crowdsensing (ACS) enabled by a range of technologies such as decentralized autonomous organizations and operations, large language models, and human-oriented operating systems. Specifically, we explain what ACS is and explore its distinctive features in comparison to traditional crowdsensing. Moreover, we present the ``6A-goal" of ACS and propose potential avenues for future research.
\end{abstract} 


\input{docs/1_Introduction}
\input{docs/2_architecture}
\input{docs/3_6A}
\input{docs/4_challenges}
\input{docs/5_conclusion}

\section*{Acknowledgment}

The authors would like to thank Yuanyuan Chen, Yonglin Tian, and Yutong Wang from Laboratory for Management and Control of Complex Systems, Institute of Automation, Chinese Academy of Sciences for attending the discussion.

\bibliographystyle{IEEEtran}
\bibliography{ref}

\section*{About the author}
\textbf{Wansen Wu} received the B.E. degrees in simulation engineering in 2018 from the National University of Defense Technology, Changsha, China, where he is currently working toward the Ph.D. degree with the College of Systems Engineering. His research interests include multi-modal learning, multi-modal large language models and mobile crowdsensing.

~\\

\textbf{Weiyi Yang} received the B.E. degree in simulation engineering from National University of Defense Technology (NUDT), Changsha, China in 2019. She is currently a Ph.D. candidate at College of Systems Engineering, NUDT, China. Her current research interests include artificial intelligence and metaheuristics, game theory approaches, and satellite scheduling problems.

~\\

\textbf{Juanjuan Li (Member, IEEE) } received the M.E. degree in economics from the Renmin University of China, Beijing, China, in 2010, and the Ph.D. degree in philosophy from the Beijing Institute of Technology, Beijing, in 2021. She is an Associate Professor with the State Key Laboratory for Management and Control of Complex Systems, Institute of Automation, Chinese Academy of Sciences, Beijing. Her research interests include blockchain, DAO, and parallel management.

~\\

\textbf{Yong Zhao} received the M.E. degree in control science and engineering from the National University of Defense Technology, Changsha, China, in 2021, where he is currently working toward the Ph.D. degree in management science and engineering. His current research focuses on crowdsensing, human-AI interaction, and intelligent transportation systems.

~\\

\textbf{Zhengqiu Zhu} received the Ph.D. degree in management science and engineering from National University of Defense Technology, Changsha, China, in 2023. He is currently a lecturer with the College of Systems Engineering, National University of Defense Technology, Changsha, China. He was also a visiting Ph.D. student in the research group of Multi-scale Networked Systems at University of Amsterdam (UvA). His research interests include mobile crowdsensing, social computing, and LLM-based AI agent.

~\\

\textbf{Bin Chen} received the B.E., M.E., and Ph.D. degrees in control science and engineering from the National University of Defense Technology, Changsha, China, in 2003, 2005 and 2010, respectively. He is currently an Associate Research Fellow with the College of Systems Engineering, National University of Defense Technology, Changsha, China. His research focuses on parallel simulation, mobile crowdsensing, and micro-task crowdsourcing.

~\\

\textbf{Sihang Qiu} received the Ph.D. degree from the Web Information Systems Group, TU Delft, Delft, The Netherlands, in 2021. He is currently an Associate Professor with the College of Systems Engineering, National University of Defense Technology, Changsha, China. His research focuses on human-centered AI, crowd computing, and conversational agents. Detailed Bio of Sihang Qiu can be found at https://sihangqiu.com/.

~\\

\textbf{Yong Peng} received the B.E., M.E., and Ph.D. degrees in control science and engineering from the National University of Defense Technology, Changsha, China, in 2004, 2006 and 2011, respectively. He is currently an associate professor at the College of Systems Engineering, National University of Defense Technology, Changsha, China. His research interests include Modeling and Simulation of complex systems, HLA/RTI, parallel and distributed simulation system, cloud-based simulation, and edge computing.

~\\

\textbf{Fei-Yue Wang (Fellow, IEEE) } received the Ph.D. degree in computer and systems engineering from the Rensselaer Polytechnic Institute, USA, in 1990. He is the State Specially Appointed Expert and the Director of the State Key Laboratory for Management and Control of Complex Systems. His current research focuses on methods and applications for parallel systems, social computing, parallel intelligence, and knowledge automation. Detailed Bio of Fei-Yue Wang can be found at www.compsys.ia.ac.cn/people/wangfeiyue.html.

\end{document}

%% file: docs/1_Introduction.tex
\section{Introduction}

\begin{figure}
    \centering
    \includegraphics[width=.85\linewidth]{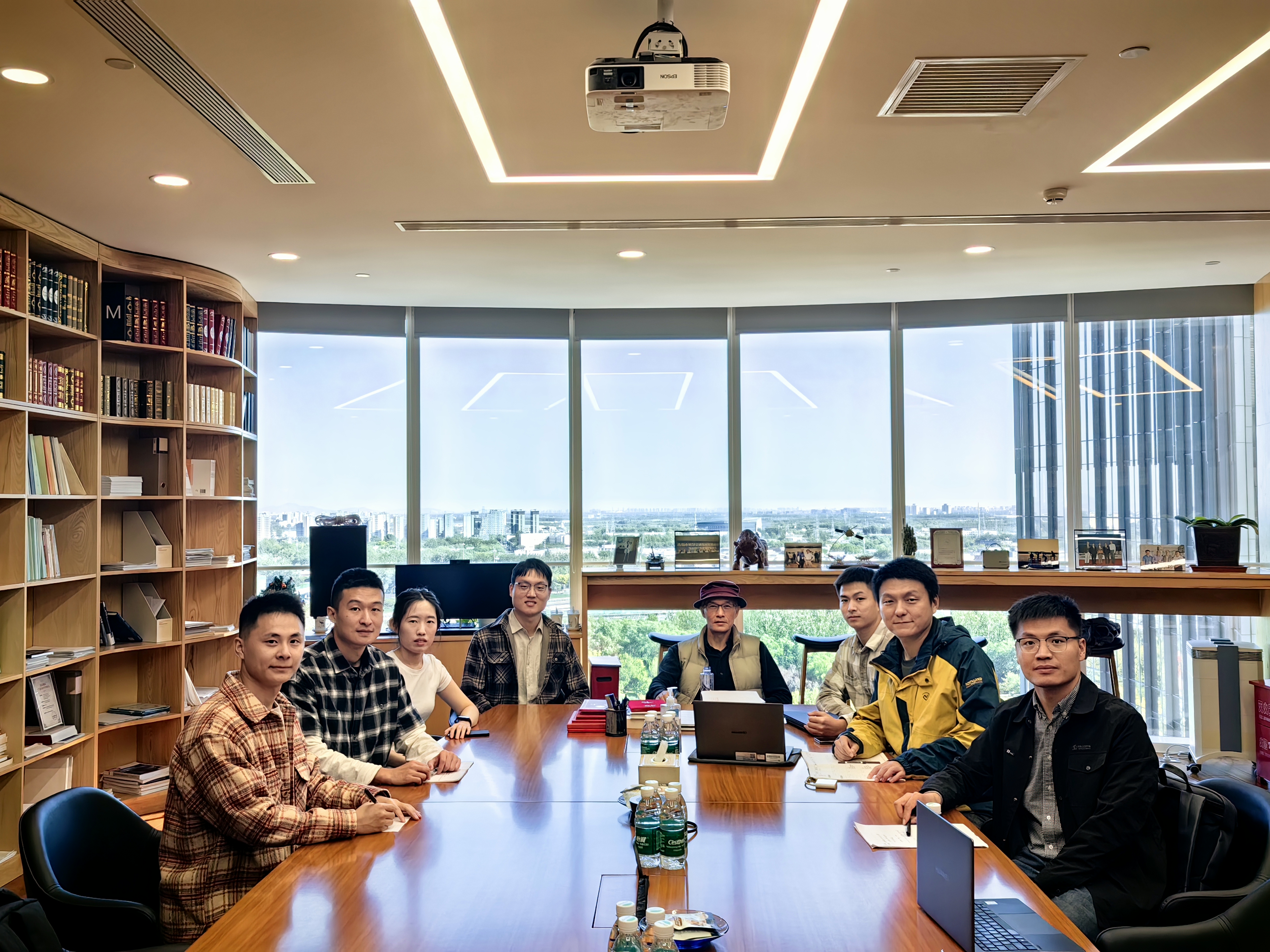}
    \caption{The latest Distributed/Decentralized Hybrid Workshop on Crowdsensing Intelligence (DHW-CSI).}
    \label{fig:seminar}
\end{figure}

\IEEEPARstart{T}{he} rapid development of the Internet and mobile computing technologies has fostered the emergence of intelligent, open, and large-scale sensing mechanisms, facilitating citizens in effectively gathering and sharing real-time information~\cite{zhu2022crowd-aided,wang2023large}. This progress enables innovative urban computing solutions to address challenges at a city-wide level~\cite{zhu2020cost}. The sensing paradigm, known as crowdsensing (CS), plays an increasingly important role in urban computing with the help of sensor-rich smartphones or wearables. In recent years, crowdsensing has made significant research advancements within the four-stage framework~\cite{zhang20144w1h} and has witnessed extensive applications across various domains.
However, the expansion of complex sensing applications has brought forth a multitude of challenges and issues for crowdsensing. These include but are not limited to: substantial human efforts, inflexible organizational structure, sluggish system response, and limited application popularization.

\begin{figure*}[htbp]
    \centering
    \includegraphics[width=.95\linewidth]{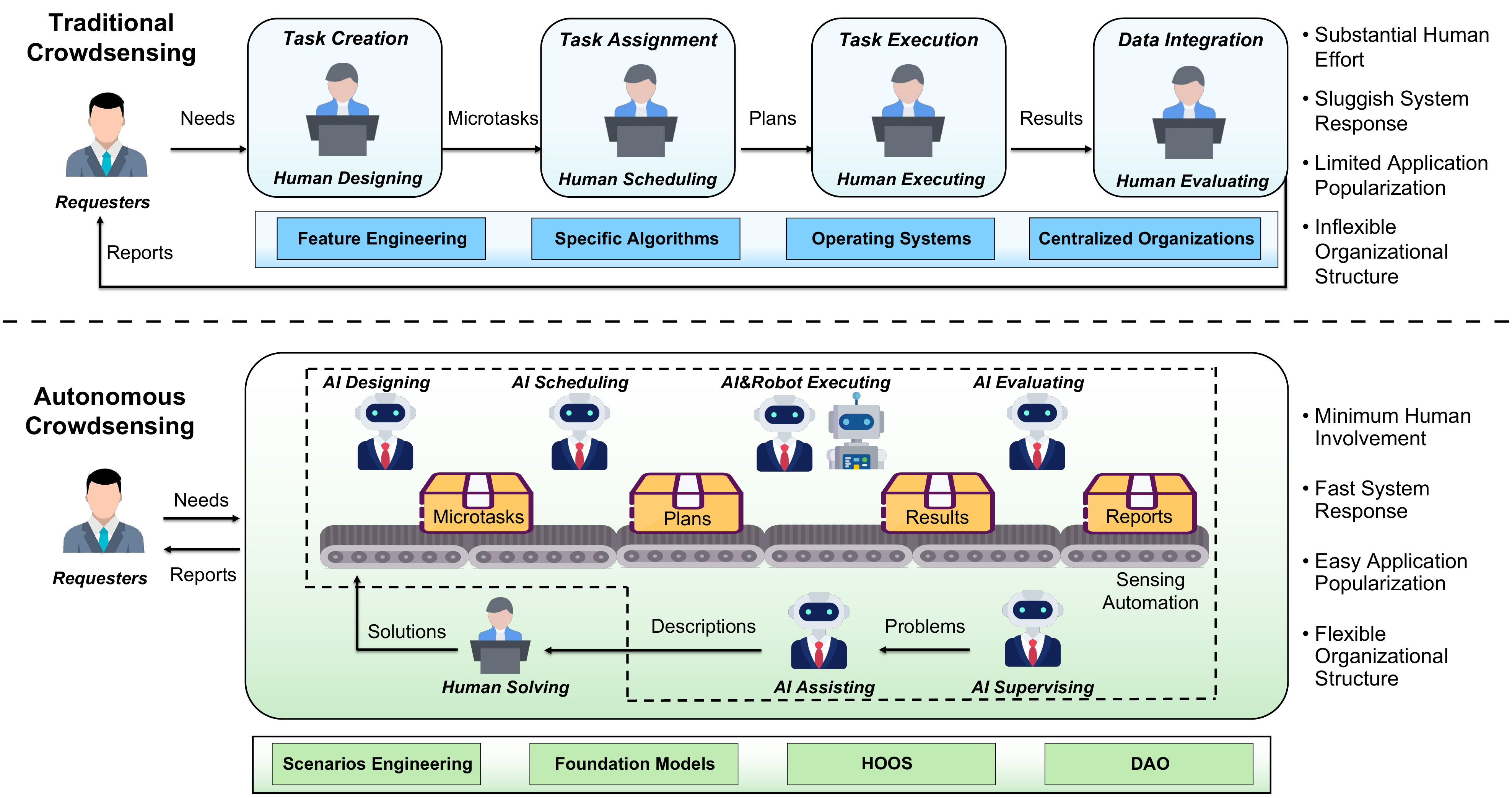}
    \caption{The traditional crowdsensing workflow v.s. the autonomous crowdsensing workflow.}
    \label{fig:teasers}
\end{figure*}

In this context, a novel generation of crowdsensing is required to effectively coordinate and schedule a diverse sensing workforce for enabling trustworthy, intelligent, interactive, collaborative, and autonomous sensing \& computing. The recent advancements in parallel intelligence~\cite{wang2023chat,zhao2023towards}, along with cutting-edge technologies such as \minew{Decentralized Autonomous Organizations and Operations (DAO)~\cite{li2022future,li2023new}} and Large Language Models (LLMs)~\cite{wang2023chatgpt0}, have inspired the proposal of a novel sensing paradigm known as Crowdsensing Intelligence (CSI)~\cite{zhu2022crowdsensing}. Through the deep integration of robotic, biological, and digital participants~\cite{zhao2023crowd,wang2023crowdsensing}, CSI leverages their diverse sensing abilities, complementary computing resources, and cross-space collaboration to build a decentralized, self-organizing, self-learning, and continuously evolving intelligent sensing and computing space. In this space, individual skills and collective cognitive ability can be enhanced to facilitate the guidance and control of the actual systems.

To achieve CSI, Professor Fei-Yue Wang and associate editor Bin Chen launched the Distributed/Decentralized Hybrid Workshop on Crowdsensing Intelligence (DHW-CSI). Our previous workshops on CSI have primarily focused on the DAO-based architecture~\cite{zhao2022decentralized,qin2022web3} and fundamental components of CSI, such as participants, methods, and stages~\cite{zhao2023crowd}. In the latest DHW-CSI, as shown in Figure \ref{fig:seminar}, we further proposed autonomous crowdsensing (ACS), a concrete form of CSI, which establishes a crowdsensing process with the goal of sensing automation. ACS organizes both professional sensing resources (e.g., satellites, IoT nodes, smart vehicles) and non-professional sensing resources (e.g., humans with smart devices) in the form of the DAO. \minew{Specifically, the utilization of smart contracts facilitates the autonomous management and execution of various tasks within crowdsensing campaigns, including the formulation of sensing plans and distribution of incentives. The specific data related to these tasks is stored in the blockchain, encompassing detailed requirements, sensing plans, and task status. In the unified architecture of the DAO, smart contracts provide the operational mechanism and business logic, while the blockchain ensures the secure and immutable data storage.}
Moreover, by leveraging the comprehensive capabilities of LLMs in comprehension, reasoning, and content generation~\cite{wang2023does,du2023chat}, LLMs-based digital humans can largely replace human labor in traditional crowdsensing tasks, thereby minimizing the involvement of biological humans. This would facilitate ACS to quickly respond to diverse sensing data demands and lower the cognitive and execution barriers for human participants, thereby enabling crowdsensing applications to achieve a greater scale of promotion in a more efficient and effective manner.

In this letter, we first discuss in detail the difference between ACS and traditional crowdsensing, and elucidate how to achieve sensing automation. Then, Section~\ref{sec:6a} presents ``6A-goal" pursued by ACS, i.e., (1) \textbf{autonomous generation}, (2) \textbf{autonomous growth}, (3) \textbf{autonomous organization}, (4) \textbf{autonomous control}, (5) \textbf{autonomous assistance}, and (6) \textbf{autonomous verification}. Finally, future research directions are discussed and conclusions are drawn.

%% file: docs/2_architecture.tex
\section{Autonomous Crowdsensing Intelligence: A New paradigm}\label{sec:arch}

The active involvement of humans is indispensable at each stage of traditional crowdsensing, as they shoulder the majority of the workload, encompassing task design, assignment, execution, evaluation, and more~\cite{zhao2023crowd,ren2022acp,wang2023chatgpt}, as depicted in the upper half of Figure~\ref{fig:teasers}. Therefore, traditional crowdsensing campaigns often require \textbf{substantial human efforts}, which poses a heavy burden for both the crowdsensing platform and the participants, particularly when dealing with large-scale and intricate sensing demands. From the platform's perspective, the requirement of substantial human efforts at each stage leads to a limited degree of automation throughout the entire process. Due to the constrained workload and diminished interest of humans, platforms often exhibit \textbf{sluggish response times} to meet data demands. Furthermore, response times are also influenced by the inherent uncertainty associated with humans. From the perspective of participants, they are requested to possess a certain level of professional knowledge and skills to successfully accomplish various tasks, which establishes a barrier to human involvement in crowdsensing campaigns and also poses \textbf{limited application popularization} of the paradigm.
Another limitation arises in \textbf{inflexible organizational structures} where a central institution or individual typically assumes responsibility for task allocation, data collection, and result processing. If this central entity experiences a failure or cannot function properly, the entire crowdsensing system could be affected, leading to task disruption or failure in data processing. 

The recent advancements in technologies such as LLMs and DAO have made remarkable progress and achieved successful applications across various fields. Furthermore, the development of these technologies has opened up new opportunities for crowdsensing. In particular, LLMs have acquired extensive knowledge and strong generalization abilities through pre-training on vast corpora~\cite{wang2023foundation,tian2023vistagpt}. As a result, they can play a crucial role in multiple aspects of crowdsensing. Besides, DAO can serve as an organizational structure that operates through the utilization of smart contracts and blockchain technology~\cite{li2023tao}. In contrast to conventional hierarchical and centralized structures, \minew{DAO integrates diverse resources within distributed, decentralized, autonomous, automated, organized, and operational communities~\cite{qin2022web3}.} It facilitates collective decision-making and resource management among a heterogeneous workforce.
\minew{In addition, the advancement of technologies like scenarios engineering, federated intelligence, and Human-Oriented Operating Systems (HOOS) can also contribute to the transformation of traditional crowdsensing. Specifically, scenarios engineering aims to shape AI systems into a form that is more relevant to the underlying scenario that will be learned and tested~\cite{li2022features}. It involves integrating application scenarios and operations within a certain temporal and spatial range to establish trustworthy and interpretable foundation models and achieves the "6S" goal~\cite{wang2023guest}. HOOS~\cite{shen2022parallel,li2022future} represents an advanced version of management and computer operating systems designed  to facilitate seamless communication and interaction between biological humans and their digital or robotic counterparts. It can effectively mitigates the burdens associated with tasks, both physical and cognitive, on individuals.}

These new technologies have facilitated the emergence of a highly autonomous and decentralized crowdsensing paradigm, resulting in the development of a new generation of crowdsensing and the establishment of crowdsensing intelligence.  
In this paper, we propose a novel Autonomous Crowdsensing paradigm to overcome the limitations associated with traditional approaches by leveraging these aforementioned technologies, as shown in Figure~\ref{fig:teasers}. To address the high demand for human involvement and mitigate sluggish system response associated with traditional crowdsensing, we believe that a multitude of tasks in crowdsensing can be effectively accomplished through agents based on LLMs, as the current generation of autonomous agents built upon LLMs has already achieved an initial level of human-like intelligence \cite{2308.11432}. Therefore, the implementation of LLMs allows for the replacement or assistance of humans in various stages of crowdsensing, thereby facilitating the automation of crowdsensing campaigns. This encompasses autonomous task design, plan scheduling, task execution, and more. As a result, the response times of crowdsensing campaigns can be significantly reduced. Additionally, humans can obtain the assistance of LLMs-based agents via HOOS, which significantly mitigates the professional quality demand for human participation in crowdsensing, facilitating widespread participation of amateurish individuals at a reduced cost. Humans primarily handle unexpected situations or complex problems within the autonomous process, resulting in a substantial reduction in workload compared to the traditional workflow. This is crucial for promoting the paradigm of intelligent crowdsensing.

\minew{DAO facilitates the formation of autonomous crowdsensing communities by assembling individuals who share common objectives, driven by specific mechanisms for value creation and incentive distribution~\cite{li2023new}. By applying DAO, ACS achieves collaboration and interaction among diverse sensing resources in a distributed organizational structure. Through the utilization of pre-programmed smart contracts, the automation and independent execution of crowdsensing campaigns can be achieved, eliminating the reliance on third-party service platforms. Moreover, DAO empowers its members with decentralized decision-making power through a democratic voting mechanism, thereby enhancing the resilience of crowdsensing activities and organizational structures. Building upon DAO, the organizational structure of ACS will further advance towards TAO (TRUE autonomous organizations and operations), characterized by being Trustable, Reliable, Usable, and Efficient/Effective~\cite{li2023dao}. TAO distinguishes itself from DAO by emphasizing power decentralization, avoiding token centralization, fostering open value systems, and facilitating integration with AI~\cite{li2023tao2}, and thereby addressing the challenges faced by DAO in practice, such as token dependency, and decision-making inertia.}

\begin{figure}[htbp]
    \centering
    \includegraphics[width=.99\linewidth]{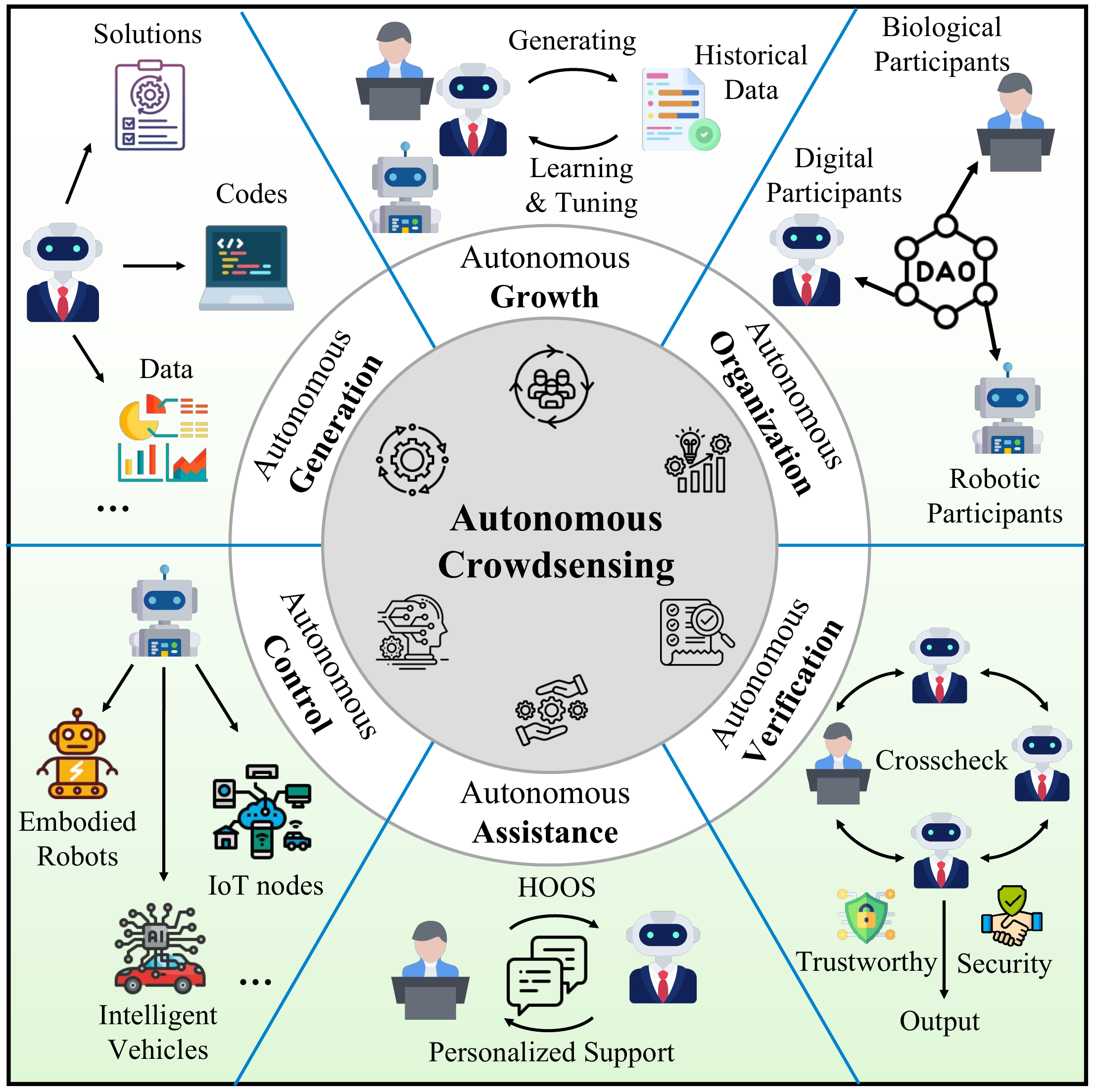}
    \caption{``6A-goal" of ACS.}
    \label{fig:6A}
\end{figure}

%% file: docs/3_6A.tex
\section{``6A-Goal" of ACS}\label{sec:6a}
Autonomous crowdsensing aims to achieve 6 critical goals, as shown in Figure~\ref{fig:6A}. 
The first three goals, i.e., autonomous generation, growth, and organization, are designed to accomplish high-level goals throughout the entire ACS workflow. In contrast, the last three practical goals primarily address the challenges encountered by three different kinds of participants~\cite{zhao2023crowd}, i.e., autonomous control for robotic participants, autonomous assistance for biological participants, and autonomous verification for digital participants. It is important to note that all the high-level goals have the potential to contribute to the three practical goals, resulting in the successful execution of the ACS workflow.

\textbf{\textit{Autonomous Generation.}}
In autonomous crowdsensing, LLMs can be leveraged to automate the generation of solutions, code, and data. For instance, LLMs are capable of effectively designing microtasks of crowdsensing according to human requests, and subsequently scheduling and optimizing these microtasks~\cite{song2023llm}. Moreover, LLMs can automatically generate code snippets or function structures based on task descriptions or requirements~\cite{imai2022github}. Relying on LLMs for autonomous generation, most of the activities in crowdsensing campaigns can be implemented automatically, eliminating the need for manual generation. This way has many benefits, such as improved efficiency and reduced reliance on manpower.

\textbf{\textit{Autonomous Growth.}}
Autonomous crowdsensing systems have the capability to utilize historical data and feedback generated during automated processes to iteratively facilitate the evolution of participants, especially for LLMs-based digital participants. This iterative process enables the models to achieve better performance and adapt to more complex tasks over time. To achieve this, these systems can leverage incremental learning algorithms~\cite{barmann2023incremental}, which continuously receive new data and update the model incrementally. By focusing on local adjustments and optimizations based on the existing LLMs, it circumvents the necessity of retraining the entire model. Furthermore, these systems can autonomously select and integrate appropriate models and algorithms based on task requirements and data characteristics, aiming to achieve better performance and adaptability. 

\textbf{\textit{Autonomous Organization.}}
\minew{Beyond the utilization of blockchain and smart contract technology, DAO is well-suited for enabling the autonomous organization of ACS due to its decentralized network governance, loose-coupled and scalable organizational structure, as well as inclusiveness in decision-making processes~\cite{yao2023towards}. Apart from the advantages of high-level transparency and traceability, these benefits promote a more inclusive and innovative environment for crowdsensing activities. The modular design of DAO also enables the integration of diverse workforces and the seamless expansion of crowdsensing network, promoting its scalability and adaptability to evolving requirements and dynamic environments.}

\textbf{\textit{Autonomous Control.}}
The autonomous control of autonomous crowdsensing systems empowers various robotic participants, embodied agents, IoT nodes, and other hardware facilities to effectively execute a diverse range of crowdsensing tasks. In these autonomous systems, LLMs act as schedulers, invoking various algorithms and tools similar to the concepts presented in AutoGPT~\cite{autogpt} and ToolFormer~\cite{schick2023toolformer}. For instance, the LLM can take into consideration various factors, including the spatial distribution, capabilities, and availability of sensing nodes, as well as the geographical location and temporal constraints associated with the tasks at hand.  By seamlessly integrating sensing, decision-making, and execution capabilities, the system can effectively govern and coordinate the actions of participants to achieve optimal task execution.

\textbf{\textit{Autonomous Assistance.}}
The autonomous crowdsensing systems can provide real-time and personalized support through the HOOS to various biological participants, tailored to their specific needs, skill levels, characteristics, and habits.  This assistance aims to facilitate the completion of diverse tasks effectively.  By incorporating personalized recommendation systems~\cite{gao2023chat} and assistive tools, the system can deliver customized support and aid based on the unique characteristics and requirements of each participant. For those amateurish workers, LLMs help reduce the learning curve by incorporating user-friendly interfaces or implementing design principles that prioritize human interaction in smart operating systems~\cite{wang2021parallel,wang2023parallel}. By automating the configuration and calibration process of device parameters for amateurish workers, accurate and consistent data collection is ensured.

\textbf{\textit{Autonomous Verification.}}
For digital participants, autonomous crowdsensing systems is capable of performing automated validation and ensuring the safety as well as feasibility of the content generated by these participants. This can be achieved through the utilization of multi-agent cooperation \cite{2308.08155} and scenarios engineering to achieve automated testing, crosscheck, and verification~\cite{weng2022large}. By employing these techniques, the system can verify the accuracy, security, and reliability of the generated schemes, data, code, and other content, thereby enhancing the credibility and effectiveness of the system. Additionally, it automates the verification and cleansing of collected data to ensure high-quality completion of sensing tasks.

%% file: docs/4_challenges.tex
\section{Challenges and Potential Avenues}

Through the introduction and discussion in the previous sections, we have proposed the ACS, a new paradigm for crowdsensing. However, some related pending issues have not been thoroughly investigated and are still being addressed. Firstly, the autonomous process heavily relies on LLMs for inference, which requires considerable hardware and software resources. In cases where sensing devices are insufficient or inaccessible, many sensing tasks may still need to be performed by humans. Additionally, the inherent biases in LLMs themselves pose a challenge to the credibility of their output~\cite{zhang2023siren}. Ensuring the trustworthiness~\cite{liu2023trustworthy} of the content generated by LLMs and mitigating bias~\cite{yeh2023evaluating} are pressing issues that need to be addressed. Lastly, the potential leakage of user privacy information through induced attacks by adversaries~\cite{duan2023privacy,helbling2023llm} is a topic that requires careful research in the future.



%% file: docs/5_conclusion.tex
\section{Conclusion}
In this letter, we present the discussion results of our latest DHW-CSI, with a particular focus on the intensional aspects and inherent characteristics pertaining to the proposed autonomous crowdsensing. With the advancement and implementation of DAO, LLMs, Scenarios Engineering, Federated Intelligence, HOOS, and other associated technologies, the organization form of various activities in crowdsensing will be reshaped, giving rise to a more autonomous, intelligent, and human-centered sensing paradigm. ACS facilitates interactions among a diverse workforce, resulting in a more user-friendly and effective utilization of diverse sensing resources for task comprehension and accomplishment. The potential impact of ACS on various domains in the future and its widespread adoption should not be underestimated. We will continue DHW-CSI in the near future. Welcome to participate, and any suggestions or proposals are greatly appreciated.